\documentclass[a4paper,11pt]{article}
\usepackage{pos}
\usepackage{physics}
\newcommand{\bonn}{
    Helmholtz-Institut f\"{u}r Strahlen- und Kernphysik,
    Rheinische Friedrich-Wilhelms-Universit\"{a}t Bonn, 53115 Bonn, Germany
}

\newcommand{\ias}{
    Institute for Advanced Simulation,
    Forschungszentrum J\"{u}lich, 54245 J\"{u}lich, Germany
}

\newcommand{\casa}{
    Center for Advanced Simulation and Analytics,
    Forschungszentrum Jülich, 52425 J\"{u}lich, Germany
}

\newcommand{\bielefeld}{
    Institut für Theoretische Physik, 
    Universität Regensburg, 93040 Regensburg, Germany
}

\title{Exploring Group Convolutional Networks for Sign Problem Mitigation via Contour Deformation}
\ShortTitle{Exploring Group Convolutional Networks for Sign Problem Mitigation}

\author*[a,b,c]{Christoph Gäntgen}
\author[a,c]{Thomas Luu}
\author[d]{Marcel Rodekamp}

\affiliation[a]{\ias}
\affiliation[b]{\casa}
\affiliation[c]{\bonn}
\affiliation[d]{\bielefeld}

\emailAdd{c.gaentgen@fz-juelich.de}

\abstract{
The sign problem that arises in Hybrid Monte Carlo calculations can be mitigated by deforming the integration manifold. 
While simple transformations are highly efficient for simulation, their efficacy systematically decreases with decreasing temperature and increasing interaction. Machine learning models have demonstrated the ability to push further, but require additional computational effort and upfront training.
While neural networks possess the capacity to learn physical symmetries through proper training, there
are anticipated advantages associated with encoding them into the network’s structure. These include enhanced accuracy, accelerated training, and improved stability. The objective of the present
study is twofold. First, we investigate the benefits of group convolutional models in comparison to
fully connected networks, with a specific focus on the effects on the sign problem and on computational aspects. Second, we examine their capabilities for transfer learning, demonstrating the ability
to further reduce training cost.  We perform our investigations on the Hubbard model on select low-dimensional systems.
}

\FullConference{The 41st International Symposium on Lattice Field Theory (LATTICE2024)\\
28 July - 3 August 2024\\
Liverpool, UK\\}

\begin{document}
\maketitle

\section{Introduction}
The Hubbard model, introduced in \cite{Hubbard1959}, is a very well known condensed matter model used to describe electronic properties of metals. By combining tight binding with onsite interaction on a fixed spatial lattice, it creates a system of strongly correlated electrons. 
\begin{equation}\label{eq:HubbardH}
	H = -\sum_{ x,y}\kappa_{x,y}\left(a^\dagger_{x\uparrow}a_{y\uparrow}^{} + a^\dagger_{x\downarrow}a_{y\downarrow}^{}\right) - \frac{U}{2}\sum_{x}{\left(n_{x\uparrow}-n_{x\downarrow}\right)}^2 - \mu \sum_x (n_{x\uparrow} + n_{x\downarrow})
\end{equation}
Equation \ref{eq:HubbardH} consists of the tight-binding term with the hopping matrix $\kappa_{x,y}$, the interaction term with coupling parameter $U$ and an optional chemical potential $\mu$. Here $a^{}$, $a^\dagger$ and $n$ are the electron-annihilation, -creation and -number operators acting on a specified spatial site and spin orientation.
This hamiltonian can be transformed into the action
\begin{equation}\label{eq:HubbardS}
	S=\sum_{x,t}\frac{\phi_{x,t}^2}{2\tilde{U}}-\log\det(M[\phi,\tilde{\kappa},\tilde{\mu}]M[-\phi,-\tilde{\kappa},-\tilde{\mu}]) \in \mathbb{C}
\end{equation}
by Hubbard-Stratonovich transformation \cite{Stratonovich,Luu2015}. Here the auxiliary scalar field $\phi$ is introduced alongside the fermion matrix $M$. The integral over imaginary time is discretized into $N_t$ slices. As a result the physical parameters are rescaled ($\tilde{U}=U\times\delta$) by the temporal lattice spacing $\delta=\beta/N_t$, which depends on the inverse temperature $\beta$.
The fermionic part of eq.~\ref{eq:HubbardS} is generally complex valued with the exception of bipartite lattices at $\mu=0$.
Therefore expectation values obtained via the typical path integral formalism require additional care,
\begin{equation}\label{eq:expectationValue}
	\expval{\hat{O}} =  \frac{1}{\mathcal{Z}}\int \mathcal{D}\phi\,  \hat{O}\left[ \phi \right]e^{-S\left[\phi_n\right]}\approx \frac{1}{N}\sum_{n=0}^N \hat{O}\left[ \phi_n \right]\ \text{with}\ \phi_n\sim e^{-S\left[\phi_n\right]}/\mathcal{Z}
\end{equation}
As indicated in eq.~\ref{eq:expectationValue} the standard procedure is to stochastically approximate the high dimensional integral with field configurations $\phi_n$ sampled from a probability distribution $e^{-S\left[\phi_n\right]}/\mathcal{Z}$. 
For complex valued actions reweighting is required to correct the otherwise complex probability distribution,
\begin{equation}\label{eq:reweighting}
	\expval{\hat{O}} = \frac{
		\expval{\hat{O} e^{-\mathrm{i}S_I} }_R} {\expval{e^{-\mathrm{i}S_I}}_{R} }
	\approx
	\frac{\sum_{n=0}^N \hat{O}\left[ \phi_n \right] e^{-\mathrm{i}S_I\left[\phi_n\right]}
	}
	{
		\sum_{n=0}^N  e^{-\mathrm{i}S_I\left[\phi_n\right]}
	}\ .
\end{equation}
The complex phase is included in the observable and the expectation value is sampled with respect to the real part of the action. Though formally exact, the issue arises numerically from limited statistics. When the average phase is small the expectation value is more susceptible to statistical fluctuations.
The sign problem refers to the slowed convergence of stochastic estimates due to the cancellation of complex phases/signs in the denominator \cite{Loh1990}. 
The absolute value of this denominator is used as a measure for the severity of the sign problem, which we call the \emph{statistical power} $\Sigma=|\expval{e^{-\mathrm{i}S_I}}_{R}|$.

The separation of real and imaginary parts introduces non-holomorphicity in the RHS of eq.~\ref{eq:reweighting}, meaning while the observable is a fixed value, the expectation values in the numerator and denominator are contour dependent. This contour deformation dependence provides a possible way to reduce the sign problem.
In \cite{Wynen2020, Alexandru:2018ddf,Cristoforetti:2013wha,Detmold:2020ncp,Rodekamp2022} contour deformation was applied successfully, making otherwise infeasible calculations due to the sign problem possible. Many of these works were facilitated by machine learning methods. 
This work explores the proper utilization of physical symmetries by directly designing the symmetries within the network.
We train the network on an approximation of the contributing Lefschetz thimbles \cite{Lefschetz1921}, which is reached by integrating holomorphic flow equations from randomly sampled $\phi$ for finite flow time \cite{Alexandru2016}. The network is trained to parametrize the imaginary part of the resulting manifold from the real part,
\begin{equation}
	\phi^* = \phi_R + i \mathcal{N}(\phi_R)\ .
\end{equation}\label{eq:trafo}
Here $\mathcal{N}$ stands for the neural network.
The symmetries of our desired manifold follow from the properties of eq.~\ref{eq:HubbardS}:
\begin{itemize}
	\item spatial geometry (translation, rotation, mirroring)
	\item temporal (translation, reversal)
	\item complex conjugate (real valued sign flip)\ .
\end{itemize}
Thus the ideal network is invariant under real valued sign flips and equivariant under the proper permutations of field components.
The lattices we investigate in this work are an equilateral triangle and a square, combined with $N_t=16$ time slices. In all cases we will consider the complex conjugation and temporal translation. On top of that we look at the $120^\circ$ rotations of the triangle, resulting in a $3\times16\times2$ fold simplification.
For the square we consider all spatial symmetries (except for the translation as it is redundant in this case), given by one mirror axis and $90^\circ$ rotations. Resulting in up to $2\times4\times16\times2$ fold symmetry. The temporal reversal is not used here for practical reasons. 

\section{Group Equivariant Convolutional Neural Networks}
For the translations of a square lattice a standard convolutionl neural network (CNN) would suffice, but for more general symmetries their functionality has to be expanded. Cohen and Welling introduced exactly this expanded functionality, the group equivariant convolutional neural network (G-CNN), in \cite{Cohen2016}. They applied it to rotated images from CIFAR10 and MNIST achieving significantly better results than comparable CNNs of that time. With the additional symmetries accounted for, their G-CNNs reached a higher degree of weight sharing and as a consequence increased expressive capacity.
For a detailed explanation we refer the reader to the original publication \cite{Cohen2016}.   Here we only provide a brief overview.
The most salient requirement is that the convolution operation commutes with the group symmetry transformations, thus ensuring \emph{equivariance}. 
Combined sequentially with other equivariant operations or layers, a fully group convolutional neural network can be created. 
Equivariance is a property similar to invariance, when the input gets transformed, the output transforms accordingly.

Within the network, kernels (filters) are transformed under the group $G$, resulting in a stack of feature maps. Acting on the input with a transformed feature map is equivalent to acting on the inversely transformed input with the original feature map. Each element of the stack is associated with a group element, e.g. the $C4$ group creates a stack associated to $e,r,r^2$ and $r^3$. This does not create additional parameters, each transformation shares the original parameters. The principle is the same as a conventional convolution layer creating an output value for each \emph{translation} of the kernel.

\newpage
\section{Results}
In this section we show the learning capabilities of different networks on flow generated training data. As suggested in equation \ref{eq:trafo} we use the real parts of flowed configurations as input and evaluate with a loss function, how well the network reproduced the imaginary parts. The loss values in this section refer to the mean squared error loss function evaluated on the validation set. For each network we created 24 randomly initialized samples. We find that the equivariant models are more susceptible to getting stuck in local minima during the training process, a phenomenon we had encountered before when experimenting with simple parametrizations, such as an elementwise Gaussian. To both showcase and counteract this, we show the minimum and maximum loss values out of all the samples.

\subsection{Triangle}
The triangle can be thought of as a linear grid of 3 points with periodic boundary conditions. Thus the rotations are equivalent to translation in this representation. The general $N-$D spatial square lattice combined with one temporal dimension form a $N+1$-D square grid. Those grids work with the default implementation of convolutional layers that are known for their application to images. Similar to images and videos, the spatial and temporal structure of field configurations has meaning. When using linear networks the information of which components are adjacent to each other is lost.
In this subsection we compare selected architectures of convolutional networks with a fully connected linear network.
\begin{figure}[h]
	\centering
	\includegraphics[width=0.49\linewidth]{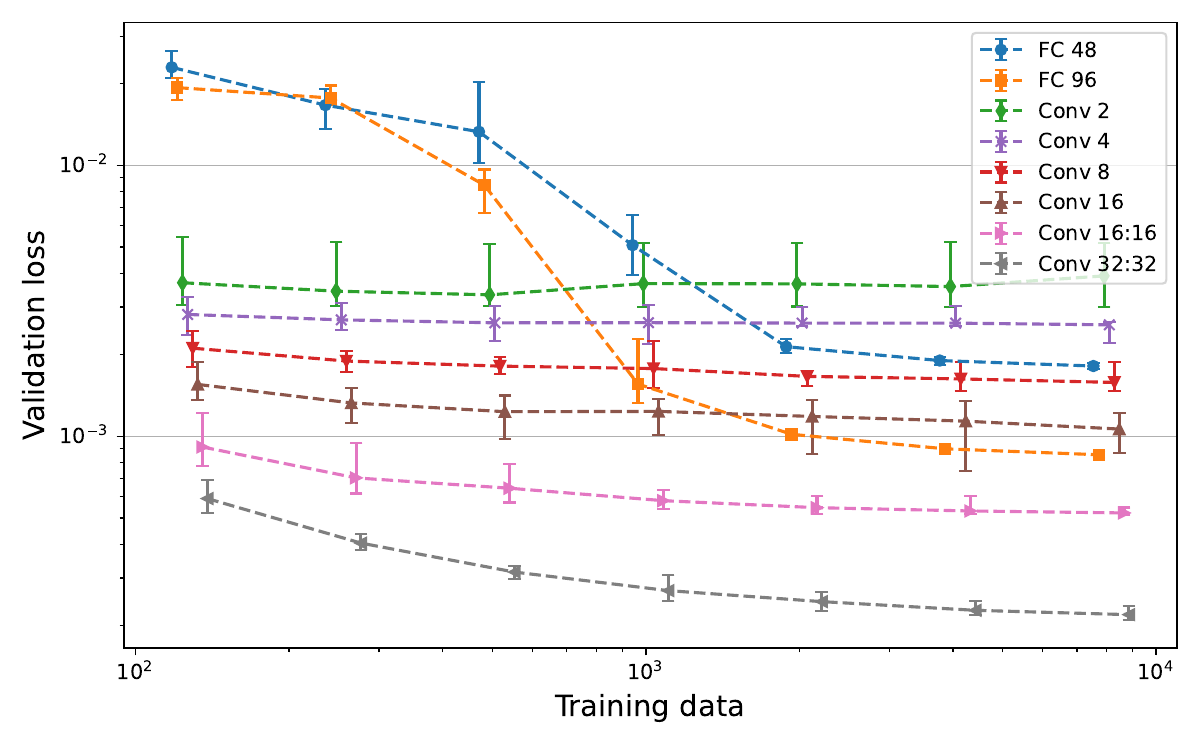}
	\includegraphics[width=0.49\linewidth]{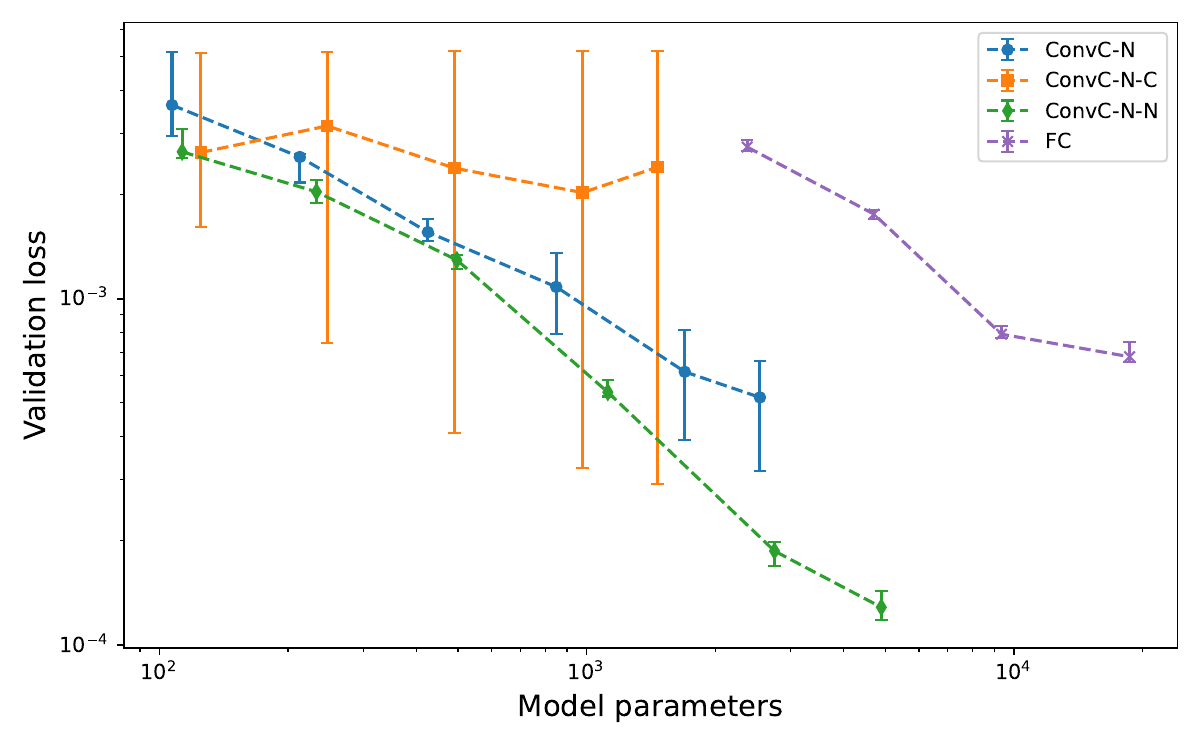}
	\caption{{\bf Left}: Final validation loss depending on amount of training configurations.  
	{\bf Right}: Final validation loss depending on tunable parameter count. The naming scheme for the labels uses \emph{C} for a convolution layer and \emph{N} for a layer in which only the channels interact with each other. Note that the error bars show the mininum and maximum results of the 24 samples, not the standard deviation.}
	\label{fig:losscomparison2}
\end{figure}
Figure \ref{fig:losscomparison2} shows a comparison of different trained models depending on available training data and size of the model. The system is a triangle at $N_t=16$, $\beta=8$ and $U=3$. Due to the odd number of sites it is non bipartite, hence it suffers from a sign problem. As expected the convolutional model has an advantage over the linear model in the case of limited training data. Also they need significantly fewer parameters to achieve the same performance.
\begin{table}[h]
	\centering
	\begin{tabular}{|c|c|c|}
		\hline
		Model & Parameters & Statistical power \\
		\hline
		Fully-Connected $96$ & $9360$ & $0.158 \pm 0.017$ \\
		\hline
		Convolutional $32$ & $1697$ & $0.117 \pm 0.021 $\\
		\hline
		Convolutional $16_{3\times3}$ & $979$ & $0.129 \pm 0.018$ \\
		\hline
		Convolutional $32:32$ & $2753$ & $0.230 \pm 0.020$ \\
		\hline
	\end{tabular}
	\caption{Statistical power evaluation of selected models.}\label{tab:SP_Conv}
\end{table}
Evaluating the statistical power on those models gives the results in table~\ref{tab:SP_Conv}.
The integers behind the model name stand for the width of the model, in case of convolutional models this is the number of channels. Two integers separated with a colon stand for two intermediate layers and the subscript stands for a second smaller convolution kernel.
For comparison we looked at two constant shifts, that we explored in depth in an earlier publication \cite{Gaentgen2023} and successfully applied in the calculation of relevant molecules \cite{Rodekamp2024}. Setting all imaginary parts to a constant, can greatly improve the sign problem. The offset that maximizes the statistical power came out to $\Sigma=0.13$. Some models exceeded this value, thereby raising optimism for challenging sign problems where the optimized shift is insufficient. Further there is a trivial shift to the plane that is tangent to the critical point of the main Lefschetz  Thimble. We can determine it analytically, it leads to $\Sigma=0.03$. We consider $\Sigma>0.1$ to be measurable with reasonable resources, showing once more that optimizing the offset expands measurable parameter space. Once this one parameter is optimized, it takes more advanced transformations to alleviate the sign problem.
\subsection{Square}
The square trivially fits a three dimensional space-time grid. We applied the \emph{C4} symmetry group, made up from rotations only. And the \emph{D4} group, containing reflections as well. As this lattice is bipartite it does not suffer from a sign problem at zero chemical potential. We induced a sign problem that is comparable to the one of the square lattice by introducing a chemical potential.
\begin{figure}[h]
	\centering
	\includegraphics[width=0.49\linewidth]{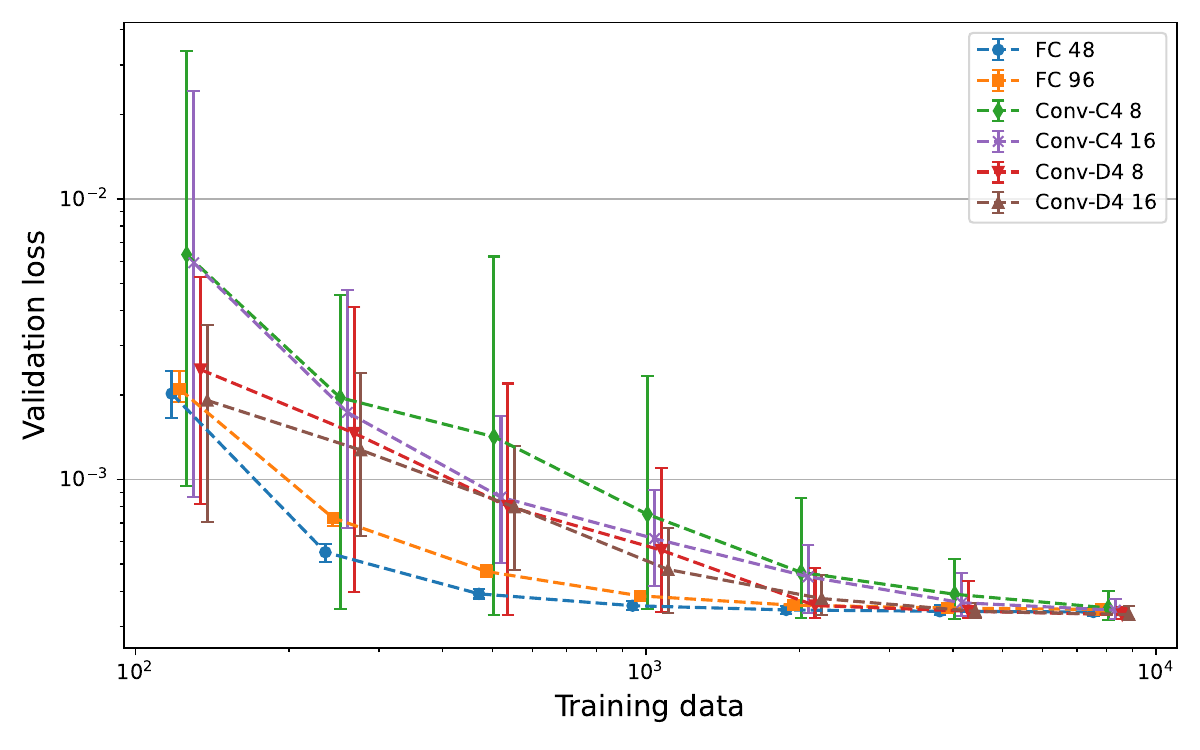}
	\includegraphics[width=0.49\linewidth]{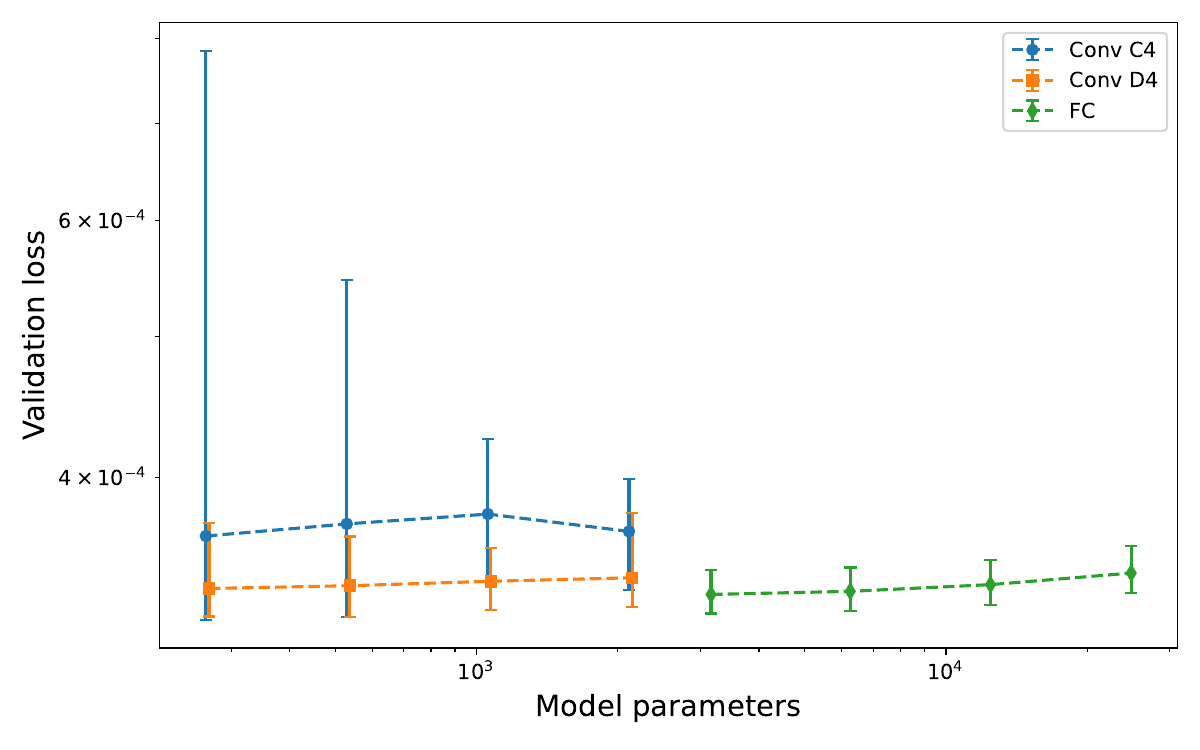}
	\caption{{\bf Left}: Final validation loss depending on amount of training configurations.  
		{\bf Right}: Final validation loss depending on tunable parameter count. Note that the error bars show the mininum and maximum results of the 24 samples, not the standard deviation.}
	\label{fig:square}
\end{figure}
Figure \ref{fig:square} shows the validation loss values for the four site square with $N_t=16$ time slices, $\beta=8$, $U=3$ and $\mu=1.5$.
Looking at the magnitude of the validation loss, it is clear that all the networks learned to reproduce the training data quite well. Unfortunately for the chosen system the generation of training data was extremely difficult, so the quality of which became a limiting factor. Still we see that the convolutional model performs as good as the fully connected network while requiring a fraction of the parameters. Further the network utilizing the \emph{D4} group with both rotations and reflections, is slightly ahead of the one using rotations only.
\section{Transfer-Learning}
A major motivation for using convolution based networks was its ability to vary the input and output dimensions. The fully connected network is created for fixed input and output dimensions, so there is no obvious way to reuse a trained model for another system size. This would be of great help for larger systems, such as graphene sheets or carbon nanotubes. Similar to the physical behavior scaling with the system size, we expected find a transferable transformation for the field components. After some testing it became clear, that this transfer between lattice dimensions would not work, as the performance of a network declines significantly when it does not see the whole input at once. Hence the networks in this paper use a convolution kernel that covers the whole input, limiting the advantages over a fully connected model to the symmetry properties only.

Nonetheless as the symmetric networks have proven to require less training data, they pose as promising candidates for classical transfer learning. Assuming that the Lefschetz thimbles (or more generally the action values $S(\phi)$) change smoothly with the model parameters $\beta$ and $U$, we can assume that the trained contour deformation for one system, is a beneficial starting point for training data generation for nearby system parameters. This introduces the cost of transforming each configuration in the process of generating training data, so we are facing a trade off between quantity and quality (presumably).
\begin{figure}[h]
	\centering
	\includegraphics[width=0.49\linewidth]{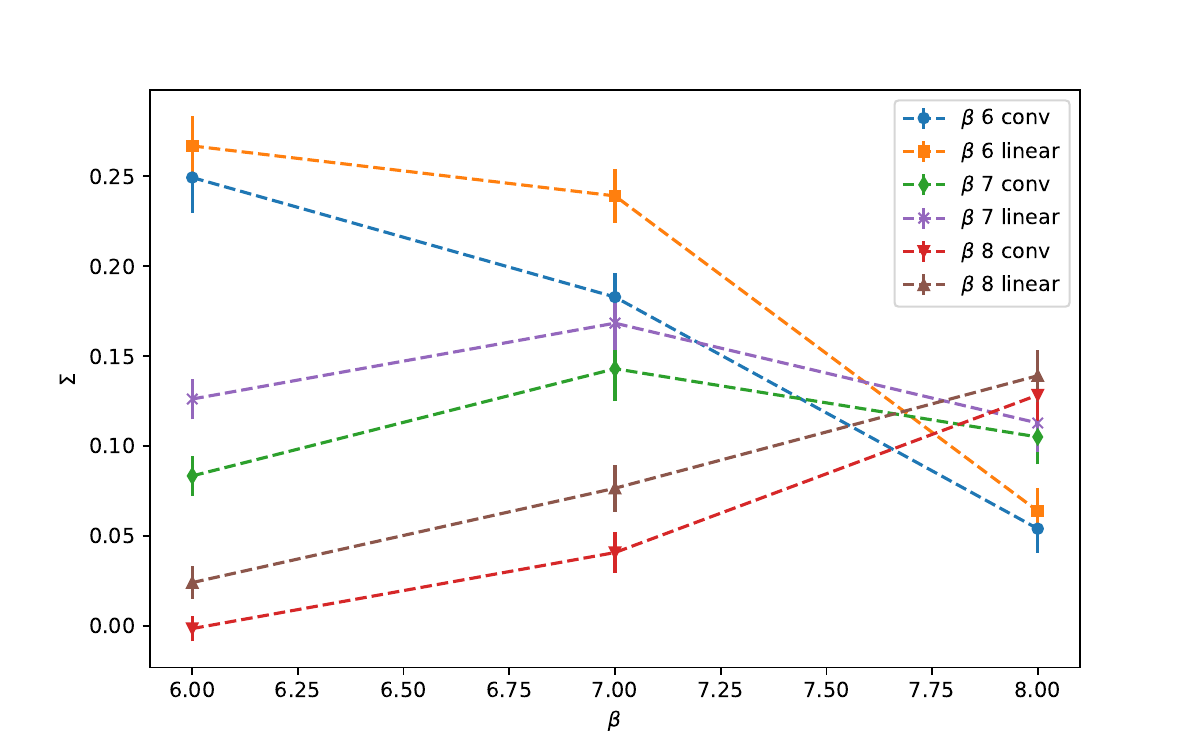}
	\includegraphics[width=0.49\linewidth]{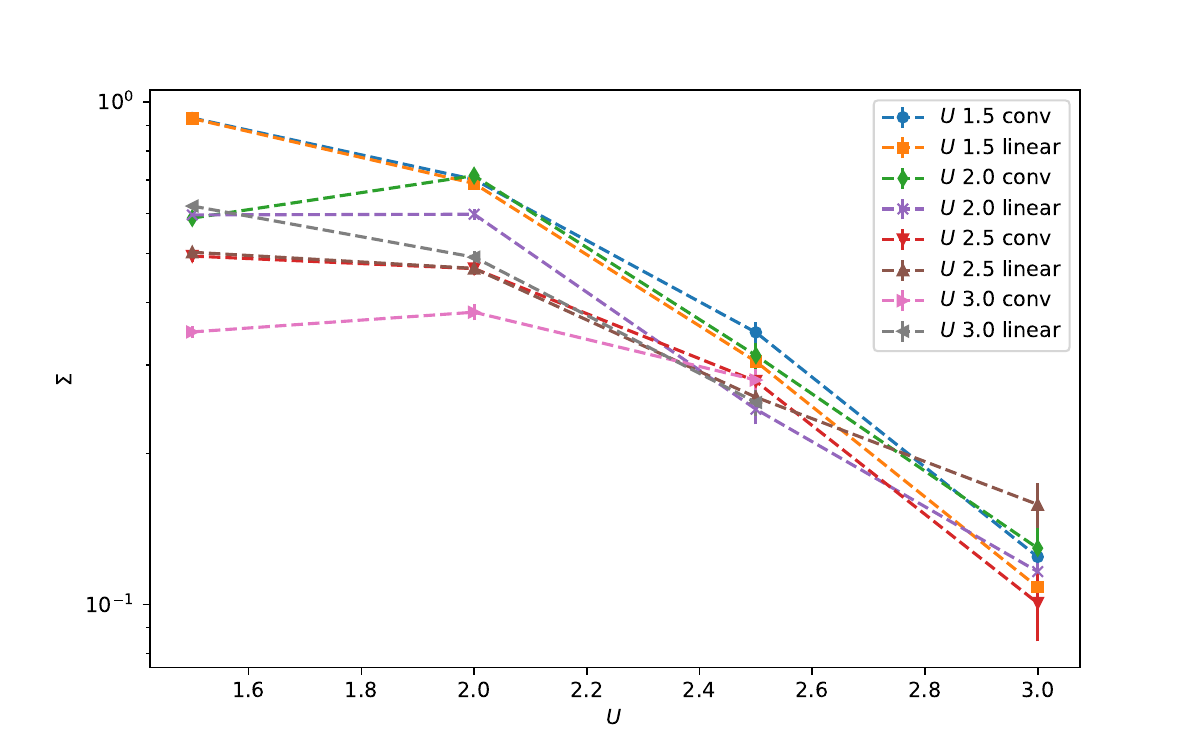}
	
	\caption{{\bf Left}: Statistical power for $\beta$ transfer learning at $U=3$. 
		{\bf Right}: Statistical power for $U$ transfer learning at $\beta=8$. The system is a triangle with $N_t=16$.
		}
	\label{fig:transfer}
\end{figure}
Figure \ref{fig:transfer} shows both linear and convolutional models that were initially trained on standard training data ($2^{15}$ samples) for one $\beta$ or $U$. Then from the learned transformation new training data was generated ($2^{12}$ samples). The model was then retrained on the new data, without resetting the already learned parameters. 
With few exceptions the transfer learned models could not keep up with the original training method. We observe though that generally the smaller transfer steps are more successful than further steps (for example $\beta=7\rightarrow8$ is better than $\beta=6\rightarrow8$), so step size clearly matters. We conclude that the applicability of this method is situational. If the task at hand is a fine analysis of a range of parameters this method can save resources and promote stability.

Further we looked at the same principle to iteratively refine one model, by repeatedly generating training data from the transformation of a trained model with holomorphic flow and retraining the model. We find that this approach has the potential to generate very good training data and models, but is unreliable. Considering the additional cost from calculating the transformations, it is preferable to start from an offset and flow further with higher precision.

\section{Summary \& Outlook}
In this proceeding we revisited a method for reducing the Hubbard model Sign problem via contour deformation that has been successfully applied in previous works \cite{Wynen2020, Gaentgen2023, Rodekamp2022,Alexandru:2018ddf,Alexandru2016,Cristoforetti:2013wha}. A neural network learns to parametrize an optimized manifold obtained by computationally expensive holomorphic flow. The original work featured fully connected neural networks which we were now able to surpass with convolutional networks. They are connected to the structure of the inputs, ensuring and leveraging the symmetries of the physical system. We observed that these \emph{informed} models excel for limited training data availability and require less tunable parameters. Further it drastically limits the risk of introducing unphysical biases, by guaranteeing that the output parametrization follows the proper symmetries. 

We investigated these models also with the expectation, they would excel at transfer learning tasks too. However we  could not confirm this assumption. Generating the training data from a related system does make sense intuitively, but introduces a layer of variability to the process of generating training data. Thus we found it had the potential to outperform our established practices, but did so unreliably. 

In the future we plan to implement the symmetries of larger carbon nano structures, such as fullerenes. The major weakness of a fully connected network was its scaling. With symmetric model architectures we hope to limit this issue, especially for systems with lots of symmetries to leverage. 

\section*{Acknowledgements}
We gratefully acknowledge the computing time granted by the JARA Vergabegremium and provided on the JARA Partition part of the supercomputer JURECA at Forschungszentrum Jülich.  Both C.G. and T.L. were support in part by the Deutsche Forschungsgemeinschaft (DFG, German Research Foundation) as part of the CRC 1639 NuMeriQS–project no. 511713970.
M.R. were supported through MKW NRW under the funding code NW21-024-A.

\newpage

\bibliographystyle{JHEP}
\bibliography{references}
\end{document}